\documentclass[%
 reprint,
 amsmath,amssymb,prl,
 aps,
]{revtex4-2}

\usepackage{graphicx}
\usepackage{dcolumn}
\usepackage{bm}
\usepackage{url}
\usepackage{amsmath}
\usepackage{amsfonts}
\usepackage{mathptmx}
\usepackage{amsthm}
\usepackage{physics}
\usepackage{bm}
\usepackage{xcolor}
\usepackage{hyperref}

\theoremstyle{definition}

\begin{document}

\preprint{APS/123-QED}

\title{Exact S-matrices for higher dimensional representations of generalized Landau-Zener Hamiltonians.}

\author{S. Malikis}
\email[]{malikis@lorentz.leidenuniv.nl}
\affiliation{Instituut-Lorentz, Universiteit Leiden, Leiden, The Netherlands}%

\author{V. Cheianov}
\affiliation{%
 Instituut-Lorentz, Universiteit Leiden, Leiden, The Netherlands}%

\date{\today}

\begin{abstract}
We explore integrable Landau-Zener-type Hamiltonians through the framework of Lie algebraic structures. By reformulating the classic two-level Landau-Zener model as a Lax equation, we show that higher-spin generalizations lead to exactly solvable scattering matrices, which can be computed efficiently for any higher-spin representation. We further extend this approach to generalized bow-tie Landau-Zener Hamiltonians, employing non-Abelian gauge fields that satisfy a zero-curvature condition to derive their scattering matrices algebraically. This method enables the systematic construction of new exactly solvable multi-level models; as a result, we present previously unknown six-dimensional and eight-dimensional Landau-Zener models.
\end{abstract}

\maketitle

\section{Introduction}

The evolution of quantum systems governed by explicitly time-dependent Hamiltonians underpins numerous contemporary phenomena and applications. Unlike time-independent scenarios, where eigenstates remain stationary up to a phase factor, time-dependent systems exhibit complex and non-trivial dynamics. This behavior is central to fields such as ultrafast spectroscopy \cite{yuen2017ultrafast,maiuri2019ultrafast}, quantum annealing \cite{hauke2020perspectives,yarkoni2022quantum}, quantum information processing \cite{nielsen2010quantum}, and quantum control \cite{d2021introduction,vandersypen2004nmr}, where understanding and manipulating time-dependent quantum behavior is essential.


However, the analytic treatment of this class of systems is an inherently difficult problem. Consider, for instance, the solution of the Schr\"{o}dinger equation. Since in general $[\hat H(t), \hat H(t')]\neq 0$, one cannot know a priori the final state of these systems, and only perturbative limits can be treated analytically \cite{landau2013quantum}. These inherent transitions between different states are known as non-adiabatic transitions \cite{nakamura2012nonadiabatic}. Therefore, one can think of this problem as a scattering problem: one starts from a basis state, and the task is to predict the scattering amplitudes to the set of possible states. In general, even the simplest case—the two-level system—cannot be solved in closed form \cite{he2024exact}.

A seminal work addressing this was the Landau-Zener model \cite{landau1932theorie,zener1932non,majorana1932atomi,stueckelberg1932two}. In particular, a two-level system is considered where the time dependence is linear in time. The system is initialized in one of the eigenstates at $t= - \infty$. The question is what the final state at $t=\infty$ will be. The full solution is known and expressed through parabolic cylinder functions; an exact formula is derived from that (the so-called "Landau-Zener formula").

In addition to the original model, many-level generalizations  have been presented that are exactly solvable. In particular, the equal slope \cite{demkov1968stationary} model, the bow-tie \cite{brundobler1993s,ostrovsky1997exact} (and its generalization in \cite{demkov2001exact}). 
Further, in \cite{yuzbashyan2018integrable,patra2015quantum,PhysRevLett.120.190402}, an algebraic treatment of the multistate Landau–Zener problem was developed by embedding the linear-in-time Hamiltonian \(H(t)=A+tB\) into families of mutually commuting operators polynomial in \(t\). This approach yields explicit zero-curvature (integrability) conditions and allows one to construct new exactly solvable LZ-type models. Moreover, larger classes of generalized models have been discovered in \cite{chernyak2018large,PhysRevB.66.205303}. An effort to classify this zoo of models was attempted in \cite{chernyak2020multitime}. Lastly, \cite{hu2024solution} presents a different solution that does not rely on any integrability condition.

These recent developments \cite{patra2015quantum,yuzbashyan2018integrable,PhysRevLett.120.190402} motivate an algebraic treatment of the problem. Using this structure, one can systematically approach the problem using tools from representation theory, making the solution more transparent. In this paper, we revisit known LZ Hamiltonians by interpreting them as elements of Lie algebras. In particular, we show that the original Landau-Zener model can be reformulated as a Lax equation and that higher-spin generalizations can be solved exactly by building on the fundamental $su(2)$ solution. Furthermore, we demonstrate how considering higher-dimensional representations naturally leads to new classes of exactly solvable Landau-Zener-type Hamiltonians, which do not fall within the previously known models.

The paper is organized as follows. In Section 1, we introduce the scattering problem in terms of the Landau-Zener framework. In Section 2, we rewrite it in Lax form and solve the higher-spin representations. In Section 3, we use the bow-tie models; by employing non-Abelian gauge fields with zero curvature with respect to the Hamiltonian, we extract new higher-dimensional Landau-Zener models.

\section{1. Scattering problem}
We start from the two-level Landau-Zener model:

\begin{equation}
    \hat H_{LZ}(t) = \Delta \hat{\sigma}_1 + t a \hat{\sigma}_3=
    \begin{pmatrix}
        a t & \Delta \\
        \Delta & -a t
    \end{pmatrix},
\end{equation}
where $\sigma_i$'s are the Pauli matrices.
Our task is solving the scattering problem, i.e. finding the amplitudes $c_i$ of the final state:
\begin{equation}
    \ket{\psi(t)}= \sum_{i=1}^2 c_i(t) \ket{i}, \quad t\rightarrow\infty,
\end{equation}
given that at $t_0 = -\infty$, the system is prepared in one of the eigenstates of $\hat H(t_0)$ $\ket{0}$.

An exact analytical solution of the Schr\"{o}dinger equation
\begin{equation}
    i \partial_t \ket{\psi(t)}=\hat H(t) \ket{\psi(t)}
\end{equation}
exists and is expressed through parabolic cylinder functions. It is worth mentioning that if the initial state is lowest energy eigenstate $\ket{0}$ of $\hat H(t_0)$, the amplitude $||c_i||$ of ending up in that state as $t \rightarrow \infty$:
\begin{equation}
    ||c_0||=\exp(-\frac{\pi \Delta^2}{2a}),\quad ||c_1||=\sqrt{1-\exp(-\frac{\pi \Delta^2}{a})}.
\end{equation}
A similar situation occurs if the system is prepared in the $\ket{1}$ state (just by interchanging the $c_0\leftrightarrow c_1$).

To facilitate the formulation, we introduce the notion of the scattering matrix:
\begin{equation}
\label{lzscater}
    S_{LZ}(\Delta,a)=\begin{pmatrix}
        u & v \\
        v & u
    \end{pmatrix}, \quad u=\text{e}^{-\frac{\pi\Delta^2}{a}},\quad  v=1-\text{e}^{-\frac{\pi\Delta^2}{a}}.
\end{equation}
Essentially, the probability of having the transition $j\rightarrow i $ (on the diabatic basis) is given by the $(S_{LZ})_{ij}$ element.

\section{2. New Solvable Models: New Scattering Matrices}
We want to find new (higher-dimensional) Landau-Zener Hamiltonians that are exactly solvable i.e., whose scattering matrix can be attained analytically. First, we exclude trivial embeddings of already known examples into higher-dimensional spaces, since they ultimately are reducible to their original. We are primarily interested in irreducible Landau-Zener Hamiltonians.


\subsection{Adjoint representations}
Given the $H_{LZ}$ we want to create new models. We are interested in exact solutions to equations of the form:
\begin{equation}
\label{general_lz}
    i\dot{\textbf{c}}=(\hat{A}+t \hat{B})\textbf{c}.
\end{equation}

A way to generate a model is the following:
We introduce the general matrix $\hat{V}=\sum_{i=1}^3 v_i(t) \hat{\sigma}_i$. We consider the Lax equation:
\begin{equation}
\label{lax}
    i\dot{\hat{V}}(t)=[\hat{V}(t),\hat{H_{LZ}}(t)]
\end{equation}
This yields a system equivalent to \eqref{general_lz}. The corresponding new Hamiltonian is:
\begin{equation}
\label{lz_ad_h}
  \hat{H}_{ad}(a,\Delta)=  \begin{pmatrix}
       2 a t & 0 & \sqrt{2} \Delta \\
       0 & -2 a t & \sqrt{2} \Delta \\
       \sqrt{2} \Delta & \sqrt{2} \Delta & 0
    \end{pmatrix}.
\end{equation}
This model was previously presented in \cite{PhysRevA.99.032112}. We provide its scattering matrix below.

\begin{equation}
\label{scaterring_ad}
    S_{ad}=\begin{pmatrix}
    u^2 & v^2 & 2 u v \\
    v^2 &  u^2 & 2 u v \\
    2 u v & 2 u v  & (1-2u^2)^2
\end{pmatrix}.
\end{equation}

In the appendix, the explicit solution is presented.

Basically, from an algebraic point of view, $\hat H_{ad}$ is the Landau-Zener Hamiltonian written in the adjoint representation of $su(2)$. From this perspective, the model might seem trivial, since it is algebraically identical to the original Landau-Zener model. However, using the same representation introduces a new model, featuring new energy levels and different level crossings.

The adjoint representation coincides with the Spin 1 representation. The formulation we have established can be used to study all the other higher-spin representations.

\subsection{Higher spin representations}
We now establish the general treatment of Landau-Zener problems with higher spin.

The $\frac{k-1}{2}$- spin Landau-Zener model is expressed through:

\begin{equation}
    \hat H_{k}= a t \hat Z_k +\Delta \hat X_k.
\end{equation}
This representation consists of k-dimensional Hermitian matrices. To find them, one should make use of an orthonormal reference basis $\{\ket{m}\}$, where $m=\{-\frac{k-1}{2},-\frac{k-1}{2}+1,\cdots,\frac{k-1}{2} \}$ with the properties:
\begin{gather}
    \hat Z_k \ket{m}=m \ket{m}, \quad (T_{k})_{\pm}\ket{m}=a_{k,m}^{\pm} \ket{m\pm 1}, \\
   (T_k)_{\pm}=\hat X_k \pm i \hat Y_k, \quad a_{k,m}^{\pm}=\sqrt{\Big(\frac{k-1}{2}\pm m\Big)\Big(\frac{k-1}{2}\mp m+1\Big)}.
\end{gather}

These expressions are irreducible k-dimensional representations of Landau-Zener Hamiltonians. In fact, one can treat all of those in a uniform way, i.e. by extracting the respective scattering matrix. 

To do so, one should have the explicit solution $\hat{V}(t)$ of Eq. \eqref{lax}.
In particular, the matrix $\hat{V}(t)$ satisfies the following properties:
\begin{enumerate}
    \item time-independent eigenvalues
    \item the eigenvectors evolve under the action of $\hat{U}_{LZ}$ (the time ordered exponential generated by the $\hat H_{LZ}$.
\end{enumerate}
    Thus, for every $\ket{c(t)}$, eigenvector of $\hat{V}(t)$:
\begin{equation}
  \exists \kappa(t): \mathbb{R} \rightarrow \mathbb{R}, \quad  \ket{c(t)}=\exp(i \kappa(t)) \hat{U}_{LZ}(t,t_0) \ket{c(t_0)}.
\end{equation}
In the context of Landau-Zener problems, if the initial eigenbasis of $\hat V(t_0)$ coincides with that of $\hat H_{ad}$, the time-evolved state is the solution of the Landau-Zener problem, as well. Thus, if $\ket{c_i}(t)$ is the eigenstate corresponding to $\lambda_i$ eigenvalue of $\hat V(t)$ and $\ket{u_j}$ is the eigenstate of the $-\hat H_{LZ}$ in $t\rightarrow \infty$, we have:

\begin{equation}
   S_{ij}=\lim_{t\rightarrow \infty }\Big|\braket{c_i(t)}{u_j}\Big|^2.
\end{equation}
Note that $H_{LZ} \to -\hat Z_k$ as $t \to -\infty$. This expression is equivalent to using the projectors onto the i-th, j-th eigenspaces, respectively. This can be efficiently computed using the matrices $\hat{V},\hat H_{LZ}$:
\begin{equation}
\label{su2generalformula}
    S_{ij}= \lim_{t\rightarrow \infty}Tr\Big(\prod_{a\neq i} \frac{\hat{V(t)}-\lambda_a \mathbb{I}}{\lambda_a-\lambda_i}\prod_{b\neq j}\frac{-\hat Z_k-\lambda_b \mathbb{I}}{\lambda_b-\lambda_j}\Big).
\end{equation}
The set of eigenvalues is $\vec{\lambda}=\{-\frac{k-1}{2},-\frac{k-1}{2}+1,\cdots,\frac{k-1}{2} \}$
Moreover, the final result depends solely on the $v_3$ component of $\hat V$. As we show in the Appendix if at $t\rightarrow - \infty$, the initial condition is chosen $v_1=v_2=0, v_3=1$:
\begin{equation}
\label{v3lab}
    v_3(t)=1-2 \text{e}^{-\frac{\pi\Delta^2}{a}}, \quad \text{as } t\rightarrow \infty.
\end{equation}

For the spin-$\frac{1}{2}$ case (i.e., the conventional Landau-Zener model):
\begin{gather}
    S_{ii}=Tr(\hat V \hat H_2-\frac{\mathbb{I}}{4})=\text{e}^{-\frac{\pi\Delta^2}{a}},  \\
    S_{ij}=Tr(\hat V \hat H_2+\frac{\mathbb{I}}{4})=1-\text{e}^{-\frac{\pi\Delta^2}{a}}, \quad i\neq j.
\end{gather}

Similarly, one can reproduce the scattering matrix (Eq. \eqref{scaterring_ad}) for the spin 1 representation.

Before concluding, we shall provide the spin - $\frac{3}{2}$ and the corresponding scattering matrix. The Hamiltonian is:
\begin{equation}
 \hat H(t)=
    \left(
\begin{array}{cccc}
 3 a t & \sqrt{3} \Delta & 0 & 0 \\
 \sqrt{3} \Delta & a t & 2\Delta  & 0 \\
 0 & 2\Delta  & -a t & \sqrt{3} \Delta \\
 0 & 0 & \sqrt{3} \Delta & -3 a t \\
\end{array}
\right).
\end{equation}

Based on the Eqs. \eqref{su2generalformula},\eqref{v3lab} the scattering matrix is
\begin{equation}
    \hat S= 
    \begin{pmatrix}
        u^3 & 3 u^2 v & 3 u v^2 & v^3 \\
        3 u^2 v & u(3u-2)^2 & (1-3u)^2 v & 3 u v^2  \\
        3u v^2 & (1-3u)^2 v &  (1-3 u)^2 v & 3 u^2 v \\
        v^3 & 3 u v^2 & 3 u^2 v & u^3
    \end{pmatrix}.
\end{equation}

Regarding the elements of the scattering of the lowest energy to the highest ones for the $\frac{N-1}{2}$ - spin representation, the following pattern is noticeable:
\begin{equation}
    S^{(N)}_{1 k} = \binom{N}{k-1} u^{N-k-1} v^{k-1}.
\end{equation}
A closed formula for all elements, unfortunately, was not possible and one must use Eq. \eqref{su2generalformula} for higher-dimensional cases.

Similar models were extracted in \cite{tchouobiap2015finite}. However, with Lax matrices, this can be done fully algebraically, simply by substituting into Eq. \eqref{lz_ad_h}.

We conclude by commenting on the relationship between the original model and the derived ones. We obtained $\tilde{H}$ from $\hat{H}$ by considering a higher-dimensional representation of the Lie algebra associated with the original group. Consequently, the new unitary operator can be expressed in terms of the original model's unitary operator by parameterizing it using the Euler angles. However, determining the Euler angles $\theta_i$ for a given $\hat{U}_0$ is a complex problem for higher-dimensional groups.  
The key message is that, while one could theoretically derive these new representations from the original ones, our approach circumvents the need for explicit Euler angle parametrization.

\section{3. Bow tie model - Zero curvature condition}
An alternative method for acquiring the scattering matrix of certain classes of Landau-Zener Hamiltonians was presented in \cite{PhysRevLett.120.190402}. In particular, we assume some Hamiltonian $\hat H= \hat H(t, \epsilon)$, where $\epsilon$ is some parameter. We introduce the operator $\hat E$, the generator of translations along $\epsilon$ direction:
\begin{equation}
    i \partial_t \ket{\psi(t,\epsilon)} = \hat H \ket{\psi(t,\epsilon)}, \quad i \partial_{\epsilon} \ket{\psi(t,\epsilon)} = \hat E \ket{\psi(t,\epsilon)}.
\end{equation}

If we find an operator $\hat{E}$ that satisfies the zero-curvature condition:
\begin{equation}
\label{zcc}
    \partial_{\epsilon} \hat{H}-\partial_t \hat{E}+i [\hat E,\hat H]=0.
\end{equation}

\begin{figure}[h]
    \centering
    \includegraphics[width=7.30cm, height=6.2cm]{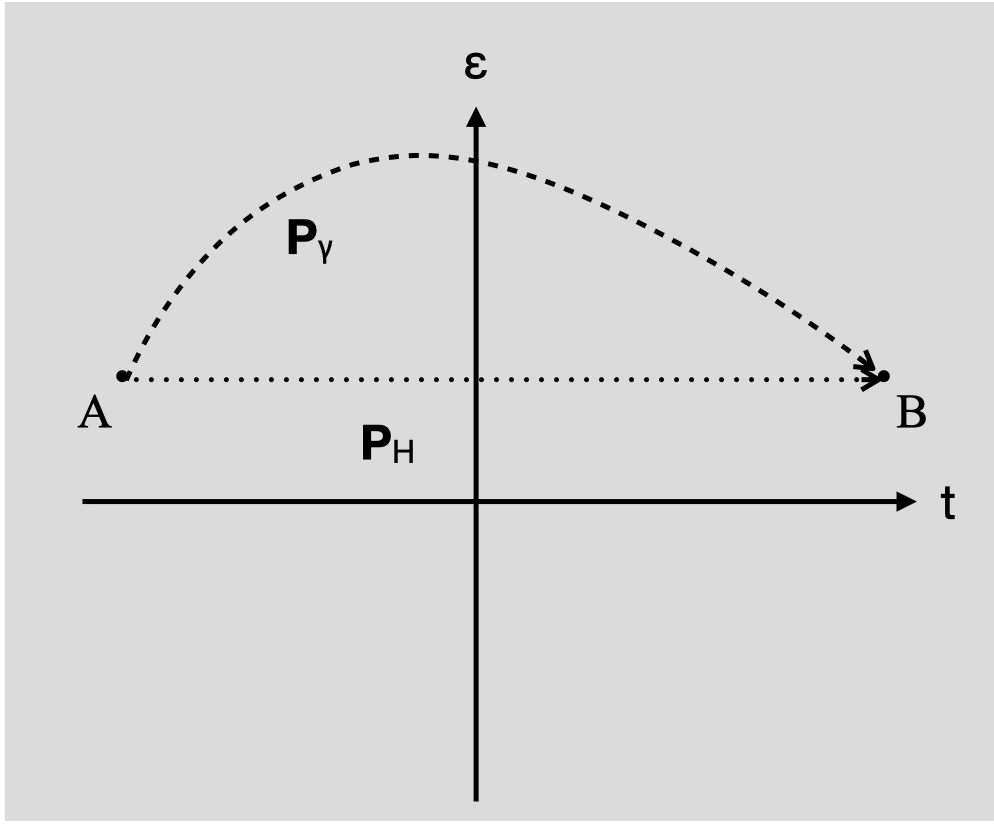}
    \caption{The evolution operator from point $A$ to point $B$ is independent from the two paths. In the regular Landau-Zener, the $\textbf{P}_H$ is employed to find the scattering matrix. With the introduction of the operator $\hat E$ one is able to deform the path arbitrarily as long as the two extreme points stay intact.}
    \label{pahtind}
\end{figure}

Given this condition and fixed endpoints of the evolution, we are free to choose any path in the $(t,\epsilon)$-plane without altering the evolution operator (see FIG. \ref{pahtind}). 

More generally, we pick a path $\mathbf{\gamma} : \mathbb{R} \rightarrow \mathbb{R}^2$, where $\mathbf{\gamma}(\tau)=t(\tau)\textbf{e}_1+\epsilon(\tau)\textbf{e}_2$. The evolution of a state along any arbitrary path $\gamma$ is generated by $\hat H_{\gamma}$:
\begin{equation}
    \hat H_{\gamma}= \frac{d t}{d \tau} \hat H+\frac{d \epsilon}{d \tau} \hat E.
\end{equation}
In the standard Landau-Zener evolution, the path is given by $t(\tau) = \tau$ and $\epsilon(\tau) = c$, a constant. 

\subsection{3-dimensional bow tie model}
We will consider the 3-level generalized bow-tie Hamiltonian:
\begin{equation}
\label{bowtie}
    H_{BT}=
    \begin{pmatrix}
        \epsilon & 0 & \Delta \\
        0 & -\epsilon & \Delta \\
        \Delta & \Delta & a t
    \end{pmatrix}.
\end{equation}

It is assumed $a>0$. This model was first introduced in \cite{demkov2001exact}. Its general solution was found by employing contour integrals.

Here we are going to do the same by using the Zero-Curvature Condition. The corresponding operator $\hat E_{BT}$ that satisfies Eq. \eqref{zcc} is given by:

\begin{equation}
    \hat E_{BT}=
     \begin{pmatrix}
        t & -\frac{\Delta^2}{a \epsilon} & -\frac{\Delta}{a} \\
       -\frac{\Delta^2}{a \epsilon} & -t & \frac{\Delta}{a} \\
        -\frac{\Delta}{a} & \frac{\Delta}{a} & \frac{\epsilon}{a} - \frac{\Delta^2}{a \epsilon}
    \end{pmatrix}.
\end{equation}

First, note that $\hat E_{BT}$ becomes singular when $\epsilon = 0$. In this case, the problem of finding the scattering matrix for $\hat H_{BT}$ becomes significantly simpler because it has a null space. As a result, for $\epsilon = 0$, $\hat H_{BT}$ reduces to an effective two-level Landau-Zener Hamiltonian.

\begin{figure}[h]
    \centering
    \includegraphics[width=7.30cm, height=6.2cm]{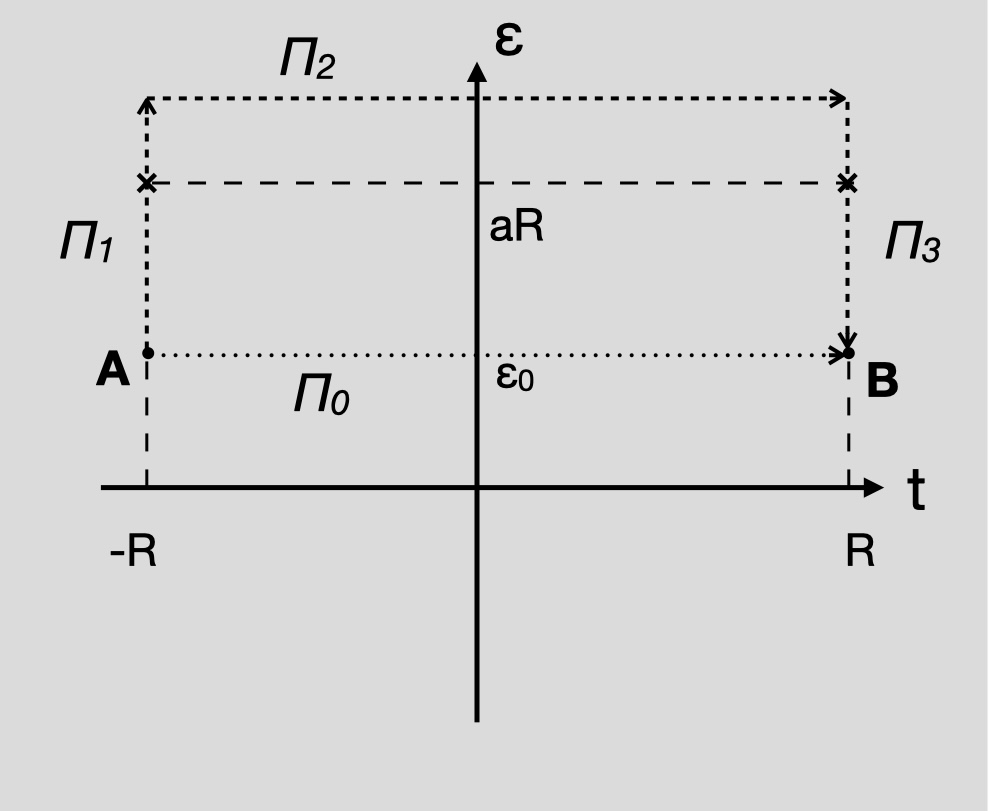}
    \caption{Path followed in the $(t,\epsilon)$ plane to compute the scattering matrix. The dotted path $\Pi_0$ corresponds to the conventional Landau–Zener evolution. Instead of this, we consider the alternate paths $\Pi_1$, $\Pi_2$, and $\Pi_3$. Level crossings occur at the marked points labeled X (at $\epsilon= a R$ and $t=\pm R$).}
    \label{pathbowtie}
\end{figure}

Let us now focus on the case $\epsilon \neq 0$. Without loss of generality, we assume $\epsilon > 0$. We deform the initial path $\Pi_0$ to $\Pi_1 + \Pi_2 + \Pi_3$ in order to find the scattering matrix (see FIG. \ref{pathbowtie}). The advantage of following these new paths is that at $t = \pm R$, where $R \to \infty$, the energy levels become well separated, with a gap proportional to $2R$, ensuring that no transition occurs. Transitions occur only at $\epsilon = aR$, where level crossings occur. At this point, the scattering reduces to a standard two-level Landau-Zener process, with the third level effectively decoupled. For instance, at $t = -R$ and $\epsilon = aR$, there is an effective interaction between levels 2 and 3.
\begin{equation}
   \hat h_1 \approx \frac{\tau}{2a} \hat \sigma_3+ \frac{\Delta}{a} \hat \sigma_1+ \mathcal{O}(R^{-1}).
\end{equation}
Similarly, at $t = R$ and $\epsilon = aR$, the transition occurs between levels 1 and 3 (with a similar effective Hamiltonian).

Therefore, the total scattering matrix is the product of the two partial scattering matrices
\begin{equation}
    S_{BT}= S_{13} S_{23},
\end{equation}
where $S_{ij}$ is a 3-dimensional matrix that acts non-trivially only on levels $i$ and $j$; the levels i,j Landau Zener scattering matrix Eq. \eqref{lzscater}. The total scattering matrix $S_{BT}$ is:
\begin{equation}
    \left(
\begin{array}{ccc}
 e^{-\frac{2 \pi  \Delta ^2}{a}} & 0 & 1-e^{-\frac{2 \pi  \Delta ^2}{a}} \\
 \left(e^{-\frac{2 \pi  \Delta ^2}{a}}-1\right)^2 & e^{-\frac{2 \pi  \Delta ^2}{a}} & e^{-\frac{4
   \pi  \Delta ^2}{a}} \left(e^{\frac{2 \pi  \Delta ^2}{a}}-1\right) \\
 e^{-\frac{4 \pi  \Delta ^2}{a}} \left(e^{\frac{2 \pi  \Delta ^2}{a}}-1\right) & 1-e^{-\frac{2
   \pi  \Delta ^2}{a}} & e^{-\frac{4 \pi  \Delta ^2}{a}} \\
\end{array}
\right).
\end{equation}
The result is independent of $\epsilon$ as long as it is a positive number; for $\epsilon<0$, one has to permute the first two columns/rows.

\subsection{N-dimensional generalized bow-tie model}

Now that we have solved the three-level case, we can generalize the method to higher-dimensional bow-tie Hamiltonians. The general form of the 
k-dimensional bow-tie Hamiltonian is

\begin{gather}
   \hat H^{(k)}=
    \begin{pmatrix}
        \hat E & \hat \Delta \\
        \hat \Delta^T &  t \hat B
    \end{pmatrix},  \quad \hat E =\text{diag}\{\epsilon,-\epsilon\}, \nonumber\\
    \hat B=\text{diag}\{a_1,\cdots,a_{k-2}\}, \quad \hat \Delta=
    \begin{pmatrix}
        \Delta_1 & \Delta_2 & \cdots & \Delta_{k-2} \\
        \Delta_1 & \Delta_2 & \cdots & \Delta_{k-2}
    \end{pmatrix}.
\end{gather}

The corresponding $E^{(k)}_{BT}$ is given by \cite{yuzbashyan2018integrable}:
\begin{gather}
    \hat E^{(k)}_{BT} = 
    \begin{pmatrix}
        \hat J & \hat D \\
        \hat D^T & \hat F
    \end{pmatrix}, \quad \hat 
    J= 
    \begin{pmatrix}
        t & h(\epsilon) \\
        h(\epsilon) & -t
    \end{pmatrix}, \\
    \hat F= \text{diag}\{h(\epsilon) + \frac{\epsilon}{a_i}\}, \quad \hat D = \begin{pmatrix}
        - \frac{\Delta_1}{a_1} & \cdots & -\frac{\Delta_{k-2}}{a_{k-2}} \\
         \frac{\Delta_1}{a_1}  & \cdots & \frac{\Delta_{k-2}}{a_{k-2}} \\
    \end{pmatrix}, \\
    h(\epsilon) = -\frac{1}{\epsilon}\sum_{i} \frac{\Delta^2_i}{a_i}.
\end{gather}

We assume that the magnitudes of the parameters \( |a_i| \) are ordered as \( |a_1| < |a_2| < \cdots < |a_{k-2}| \). Additionally, we separate the positive levels (those with \( a_i > 0 \)) into the set \(\mathbf{p}\), and the negative levels (those with \( a_i < 0 \)) into the set \(\mathbf{n}\).

Following the procedure described for the three-level case, we claim that level crossings occur at specific values of \(\epsilon\), given by \( \epsilon_i = |a_i| R \), where \(R\) is a constant (and $R\rightarrow\infty$). Starting from an initial finite value \(\epsilon_0\), the first crossing occurs at \(\epsilon_1 = |a_1| R\), then the second at \(\epsilon_2 = |a_2| R\), and so on, assuming \(\epsilon_0 > 0\).

At each value \(\epsilon_i\), the \(i\)-th level crosses with level 2 if \(i \in \mathbf{p}\) (i.e., if \(a_i\) is positive), or with level 1 if \(i \in \mathbf{n}\) (i.e., if \(a_i\) is negative). At these crossing points, the effective Hamiltonian reduces to a two-level Landau-Zener Hamiltonian.

\begin{equation}
    \hat{\tilde H}_i = \frac{\Delta_i}{a_i} \hat \sigma_1 +\frac{t}{2 a_i} \hat \sigma_3.
\end{equation}

We denote the corresponding scattering matrices by \( \tilde S_{c_i} \).

At \( t = R \), a second series of crossings occur at the same values of \(\epsilon_i\) (but in reverse order). In this case, the levels in \(\mathbf{p}\) (positive levels) will now cross with level 1, while the levels in \(\mathbf{n}\) (negative levels) will cross with level 2. The scattering matrices in this case are denoted by \( \tilde S_{d_i} \). Moreover, note that the time distance between consecutive crossings is proportional to R; the long time intervals imply that the scattering amplitudes converge to those predicted from the 2-level Landau-Zener model. Thus, the total scattering matrix takes the form:

\begin{equation}
    S=\tilde S_{d_1} \tilde S_{d_2} \cdots \tilde S_{d_{k-2}} \tilde S_{c_{k-2}} \cdots \tilde S_{c_{1}} .
\end{equation}

\section{4. Higher dimensional representations of bow-tie models}

We have derived the scattering matrix for the \(n\)-dimensional bow-tie Hamiltonian by using the zero-curvature pair \(\hat{E}\) of the Hamiltonian. Although the scattering amplitudes were already known from previous work \cite{demkov2001exact}, our approach is purely algebraic. That is, the pair \(\hat{H}(t, \epsilon)\) and \(\hat{E}(t, \epsilon)\) can be interpreted as elements of the corresponding Lie algebra. Consequently, this allows for the construction of new, exactly solvable higher-dimensional models.

It is important to note that the exact solvability of the system depends on several factors. These include the existence of the operator \(\hat{E}\), the condition that the different level crossings are well separated (which ensures factorizability) and the ability to reduce the problem to known elementary transitions, such as two-level Landau-Zener transitions, by following an equivalent path.


If we regard the three-level bow-tie model as an element of the \( su(3) \) algebra, we can have a higher non-reducible representation of $su(3)$. Then the corresponding operator \( \hat{E} \) is immediately known. In principle, one would be able to find the new scattering matrix, just by rewriting the unitary operator in Euler angles. However, even in the \( su(3) \) case, this is a formidable task. It is easier to apply the same analysis with the level crossings.

The next irreducible representation of \( su(3) \) occurs in six dimensions \cite{pfeifer2003lie}.  The Hamiltonian takes the form:
\begin{equation}
    \hat H^{(6)}_{BT}=
    \left(
\begin{array}{cccccc}
 2 \varepsilon  & 0 & 0 & \sqrt{2} \Delta  & 0 & 0 \\
 0 & 0 & 0 & \Delta  & \Delta  & 0 \\
 0 & 0 & -2 \varepsilon  & 0 & \sqrt{2} \Delta  & 0 \\
 \sqrt{2} \Delta  & \Delta  & 0 & a t+\varepsilon  & 0 & \sqrt{2} \Delta  \\
 0 & \Delta  & \sqrt{2} \Delta  & 0 & a t-\varepsilon  & \sqrt{2} \Delta  \\
 0 & 0 & 0 & \sqrt{2} \Delta  & \sqrt{2} \Delta  & 2 a t \\
\end{array}
\right).
\end{equation}

\begin{figure}[h]
    \centering
    \includegraphics[width=7.30cm, height=6.2cm]{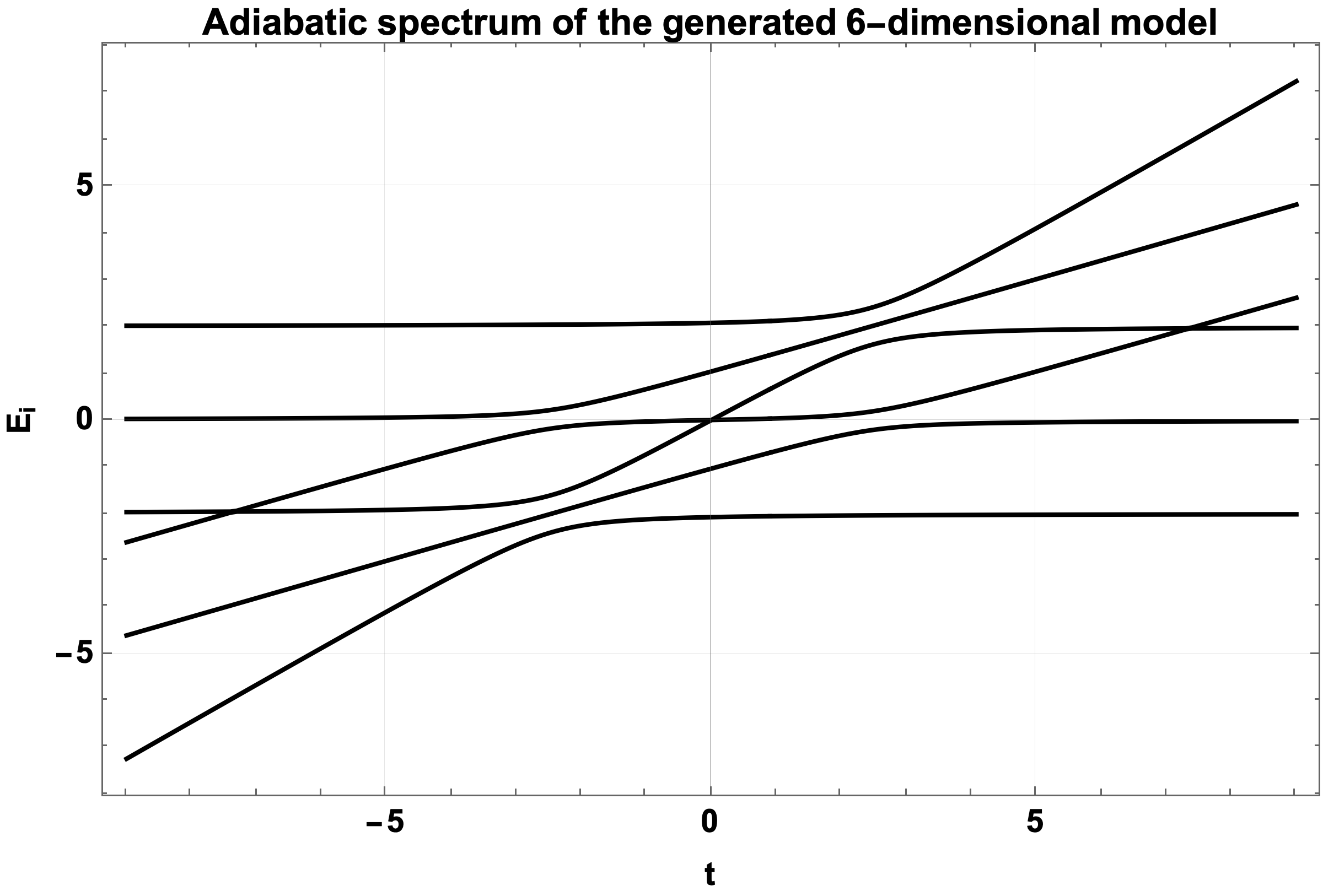}
    \caption{Adiabatic spectrum of the generated six-dimensional Hamiltonian $\hat{H}_{\mathrm{BT}}^{(6)}$ as a function of time. The parameters $\Delta=0.2$, $a=0.4$, $\epsilon=1.0$.}
    \label{adiabatic_spec}
\end{figure}

This is a new multi-level non-reducible Landau-Zener Hamiltonian whose scattering matrix can be computed exactly. The level diagram is shown in FIG. \ref{adiabatic_spec}. While the structure of this model resembles a special case of the more general $2\times3$ model proposed in \cite{chernyak2018large}, a detailed comparison of the spectra confirms that the models are distinct.

The respective $E^{(6)}_{BT}$ operator is:
\begin{gather}
    E^{(6)}_{BT}=
\left(
\begin{array}{cccccc}
 2 t & K & 0 & -\frac{\sqrt{2} \Delta }{a} & 0 & 0 \\
 K & 0 & K & \frac{\Delta }{a} & -\frac{\Delta }{a} & 0 \\
 0 & K & -2 t & 0 & \frac{\sqrt{2} \Delta }{a} & 0 \\
 -\frac{\sqrt{2} \Delta }{a} & \frac{\Delta }{a} & 0 & L+t & K & -\frac{\sqrt{2} \Delta }{a} \\
 0 & -\frac{\Delta }{b} & \frac{\sqrt{2} \Delta }{a} & K & L-t & \frac{\sqrt{2} \Delta }{a} \\
 0 & 0 & 0 & -\frac{\sqrt{2} \Delta }{a} & \frac{\sqrt{2} \Delta }{a} & 2 L \\
\end{array}
\right), \nonumber \\
K=-\frac{\sqrt{2} \Delta^2}{a \epsilon}, \quad L=-\frac{\Delta^2}{a \epsilon} +\frac{\epsilon}{a}.
\end{gather}

As outlined above, we must identify the non-trivial scatterings  for a path similar to Fig. \ref{pathbowtie}. Note that in this case we observe crossings resembling those of the three-level system in Eq.~\eqref{lz_ad_h}. In Table \ref{agg_cross}  we collect information regarding the level crossings:
\begin{table}[]
\begin{tabular}{|c|c|c|c|c|}
  \hline
   Index & Path & Point & Levels & L-Z parameters \\
  \hline
  1 & $\Pi_1$  & (-R,a R) & 2, 4 & $\Delta=\frac{\Delta}{a},\quad a=\frac{1}{2a}$ \\
  \hline
  2 & $\Pi_1$ & (-R,a R) & 6, 3, 5  & $\Delta=\frac{\sqrt{2}\Delta}{a},\quad  a=\frac{1}{a}$  \\
  \hline
  3 & $\Pi_1$ & (-R, 3aR) &  3, 4 & $\Delta=0,\quad a=\frac{1}{2a}$ \\
  \hline
  4 & $\Pi_2$ &  (0,K a R)& 2, 6 & $\Delta=0,\quad a=a$  \\
  \hline
  5 & $\Pi_3$ & (R,3bR) & 1, 5 & $\Delta=0$, $a=\frac{1}{2a}$  \\
  \hline
  6 & $\Pi_3$ & (R,bR) &  2, 5 & $\Delta=\frac{\Delta}{a} ,\quad a=\frac{1}{2a}$\\
  \hline
  7 & $\Pi_3$ & (R,bR) & 1, 6, 4 & $\Delta=\frac{\sqrt{2}\Delta}{a} \quad a=\frac{1}{a}$ \\
  \hline
\end{tabular}
\caption{The table of different crossings. In the case of the 3 level crossing that corresponds to the Hamiltonian Eq. \eqref{lz_ad_h}, the first 2 levels refer to the ones with the time dependence on their diagonal elements. The value K is some real number $K>3$.}
\label{agg_cross}
\end{table}

Notice that out of all, only crossings indexed by \(1, 2, 6, 7\) contribute non-trivially to the system's evolution. Indeed, the total scattering matrix is the product of the corresponding local transitions. This result is almost trivial at the level of unitary operators, due to the zero-curvature; however, in that case this holds on level of scattering matrix. This means that only the amplitudes of the unitary operators elements are sufficient to describe the total scattering.

We have also constructed a similar result for the original three-level generalized bow-tie model by using the adjoint representation of $su(3)$. The resulting representation is an eight-dimensional matrix, which to our knowledge is novel. The explicit form of matrices is given in Appendix II.



\section{Conclusions}
In this paper, we study the Landau-Zener (LZ) models from a Lie algebraic perspective. Our aim was twofold: first, to explore whether the scattering matrix can be obtained more easily using algebraic techniques, and second, to construct new, higher-dimensional Hamiltonians.

By adapting the Lax pair formalism to the original Landau-Zener model, we derived its adjoint representation — corresponding to a spin-1 system. This formulation enabled us to demonstrate how higher-spin representations can be systematically reduced to tractable algebraic problems, bypassing the need for full dynamical solutions or using Euler angles.

Furthermore, we extended our approach to study generalized bow-tie models in higher-dimensional representations. By employing a zero-curvature condition for gauge fields, we showed that the corresponding scattering matrices can be efficiently computed by tracking level crossings along carefully chosen deformed time evolution paths. This geometric insight allowed us to construct new classes of higher-dimensional LZ Hamiltonians, enriching the landscape of exactly solvable non-adiabatic transition models.



{\it Acknowledgments} VC gratefully acknowledges Emil Yuzbashyan for inspirational discussions. This publication is part of the project Adiabatic Protocols in Extended Quantum Systems, Project No 680-91-130, which is funded by the Dutch Research Council (NWO).

\nocite{*}

\bibliography{main.bib}

\widetext

\appendix
\begin{section}{APPENDIX I: The solution of adjoint }
The solution of Schr\"{o}dinger equation for the $H_{ad}(t)$  Eq. \eqref{lz_ad_h}:
\begin{gather}
    c_1(t)=c_1 D_{-\frac{i \Delta ^2}{2 a}}\left((1+i) \sqrt{a} t\right){}^2+c_2 D_{\frac{i \Delta ^2}{2
   a}-1}\left((-1+i) \sqrt{a} t\right) D_{-\frac{i \Delta ^2}{2 a}}\left((1+i) \sqrt{a} t\right)+c_3
   D_{\frac{i \Delta ^2}{2 a}-1}\left((-1+i) \sqrt{a} t\right){}^2, \\
   c_2(t)=-\frac{2 i a c_1 \left(D_{1-\frac{i \Delta ^2}{2 a}}\left((1+i) \sqrt{a} t\right)-(1+i) \sqrt{a} t
   D_{-\frac{i \Delta ^2}{2 a}}\left((1+i) \sqrt{a} t\right)\right){}^2}{\Delta ^2}+ \frac{2 i a c_3 D_{\frac{i \Delta ^2}{2 a}}\left((-1+i) \sqrt{a} t\right){}^2}{\Delta ^2} +\nonumber\\
   +\frac{2 a c_2 D_{\frac{i \Delta ^2}{2 a}}\left((-1+i) \sqrt{a} t\right) \left(D_{1-\frac{i \Delta
   ^2}{2 a}}\left((1+i) \sqrt{a} t\right)-(1+i) \sqrt{a} t D_{-\frac{i \Delta ^2}{2 a}}\left((1+i)
   \sqrt{a} t\right)\right)}{\Delta ^2}, \\
   c_3(t)=\frac{\sqrt[4]{-1} \sqrt{a} D_{\frac{i \Delta ^2}{2 a}-1}\left((-1+i) \sqrt{a} t\right) \left(2 c_3
   D_{\frac{i \Delta ^2}{2a}}\left((-1+i) \sqrt{a} t\right)-i c_2 D_{1-\frac{i \Delta ^2}{2
   a}}\left((1+i) \sqrt{a} t\right)\right)-2 \sqrt{2} a c_1 t D_{-\frac{i \Delta ^2}{2 a}}\left((1+i)
   \sqrt{a} t\right){}^2}{\Delta }+ \nonumber \\
   +\frac{\sqrt{a} D_{-\frac{i \Delta ^2}{2 a}}\left((1+i) \sqrt{a} t\right) \left(\sqrt[4]{-1} c_2
   D_{\frac{i \Delta ^2}{2 a}}\left((-1+i) \sqrt{a} t\right)-2 (-1)^{3/4} c_1 D_{1-\frac{i \Delta
   ^2}{2 a}}\left((1+i) \sqrt{a} t\right)-\sqrt{2} \sqrt{a} c_2 t D_{\frac{i \Delta ^2}{2
   a}-1}\left((-1+i) \sqrt{a} t\right)\right)}{\Delta }.
\end{gather}

It is interesting to study the behavior for infinite times. For that we are going to use the asymptotic behavior of $D_{\nu}(z)$. It is known that as $|z|\rightarrow \infty$ \cite{abramowitz1948handbook}:
\begin{gather}
    D_{\nu}(z)\approx \exp(-\frac{z^2}{4}) z^{\nu}\big(1+\mathcal{O}(z^{-2})), \quad   |arg(z)|<\frac{3\pi}{4}, \\
    D_{\nu}(z)\approx e^{-\frac{z^2}{4}} z^{\nu} \pm \frac{i \sqrt{2 \pi}}{\Gamma(-\nu)}e^{(-i \pi (\nu+\frac{1}{2}))} e^{\frac{z^2}{4}} z^{-\nu-1}, \quad \frac{\pi}{4}<\pm arg(z) <\frac{5\pi}{4}.
\end{gather}

Therefore the solution at $t \rightarrow - \infty$:
\begin{gather}
    c_1(t) \approx c_1 2^{-\frac{i \Delta ^2}{2 a}} e^{-\frac{3 \pi  \Delta ^2}{4 a}-i a t^2} \left(a
   t^2\right)^{-\frac{i \Delta ^2}{2 a}}, \nonumber \\
   c_2(t)\approx \frac{2^{\frac{i \Delta ^2}{2 a}} e^{-\frac{\pi  \Delta ^2}{4 a}+i a t^2} \left(a t^2\right)^{\frac{i
   \Delta ^2}{2 a}} \left(c_3 \Delta ^2 \left(-e^{\frac{\pi  \Delta ^2}{2 a}}\right) \Gamma
   \left(\frac{i \Delta ^2}{2 a}\right)^2+2 \sqrt{2 \pi } a c_2 e^{\frac{\pi  \Delta ^2}{4 a}} \Gamma
   \left(\frac{i \Delta ^2}{2 a}+1\right)+2 \pi  c_1 \Delta ^2\right)}{\Delta ^2 \Gamma \left(\frac{i
   \Delta ^2}{2 a}\right) \Gamma \left(\frac{i \Delta ^2}{2 a}+1\right)}, \nonumber \\
   c_3(t) \approx \frac{\sqrt{a} e^{-\frac{\pi  \Delta ^2}{2 a}} \left(\sqrt[4]{-1} c_2 e^{\frac{\pi  \Delta ^2}{4
   a}}+\frac{(2-2 i) \sqrt{\pi } c_1}{\Gamma \left(\frac{i \Delta ^2}{2 a}\right)}\right)}{\Delta }.
\end{gather}
For example, if we want at time $t_0$ (for $t_0 \rightarrow -\infty$) at the $(1,0,0)$ state:
\begin{equation}
   c_1= 2^{\frac{i \Delta ^2}{2 a}} e^{\frac{3 \pi  \Delta ^2}{4 a}+i a t_0^2} \left(a t_0^2\right)^{\frac{i
   \Delta ^2}{2 a}}, \quad c_2= \frac{i \sqrt{\pi } 2^{\frac{3}{2}+\frac{i \Delta ^2}{2 a}} e^{\frac{\pi  \Delta ^2}{2 a}+i a t_0^2}
   \left(a t_0^2\right)^{\frac{i \Delta ^2}{2 a}}}{\Gamma \left(\frac{i \Delta ^2}{2 a}\right)}, \quad 
   c_3=-\frac{  2^{1+\frac{i \Delta ^2}{2 a}}\pi e^{\frac{\pi  \Delta ^2}{4 a}+i a t_0^2} \left(a
   t_0^2\right)^{\frac{i \Delta ^2}{2 a}}}{\Gamma \left(\frac{i \Delta ^2}{2 a}\right)^2}.
\end{equation}
The $c_i$'s are $t_0$-dependent; however, their norms doesn't depend on  time. This happens due to the oscillatory behaviour of the solution. 

Now if one does the same process for $t\rightarrow \infty$:
\begin{gather}
    c_1(t) \approx e^{-i a t^2} \left(\frac{\sqrt{\pi } 2^{-\frac{i \Delta ^2}{2 a}} e^{-\frac{\pi  \Delta ^2}{4 a}}
   \left(a t^2\right)^{-\frac{i \Delta ^2}{2 a}} \left(c_2 2^{\frac{1}{2}+\frac{i \Delta ^2}{4 a}}
   e^{\frac{\pi  \Delta ^2}{8 a}} \Gamma \left(1-\frac{i \Delta ^2}{2 a}\right) \left((1+i) \sqrt{a}
   t\right)^{-\frac{i \Delta ^2}{2 a}} \left(a t^2\right)^{\frac{i \Delta ^2}{4 a}}+2 \sqrt{\pi }
   c_3\right)}{\Gamma \left(1-\frac{i \Delta ^2}{2 a}\right)^2}+c_1 \left((1+i) \sqrt{a}
   t\right)^{-\frac{i \Delta ^2}{a}}\right), \nonumber \\
   c_2(t)= \frac{i a c_3 2^{1+\frac{i \Delta ^2}{2 a}} e^{-\frac{3 \pi  \Delta ^2}{4 a}+i a t^2} \left(a
   t^2\right)^{\frac{i \Delta ^2}{2 a}}}{\Delta ^2}, \nonumber \\
   c_3(t)= \frac{\sqrt{a} e^{-\frac{\pi  \Delta ^2}{2 a}} \left(\sqrt[4]{-1} c_2 \Delta ^2 e^{\frac{\pi  \Delta
   ^2}{4 a}}-\frac{(4-4 i) \sqrt{\pi } a c_3}{\Gamma \left(-\frac{i \Delta ^2}{2
   a}\right)}\right)}{\Delta ^3}.
\end{gather}
 By plugging in the constants $c_i$ constants the $|c_i(t)|^2$ at $t\rightarrow + \infty$ coincides with the ones from the scattering matrix Eq. \eqref{scaterring_ad}.

Similarly one can do the same for the other linearly independent initial configurations.

\end{section}

\begin{section}{APPENDIX II: An 8-dimensional model }
The 3-level generalized bow tie model written in the adjoint representation is:
\begin{equation}
    \hat H_{ad,bt}=
\left(
\begin{array}{cccccccc}
 0 & 0 & 0 & 0 & 0 & -i \sqrt{\frac{3}{2}} \Delta  & -i \sqrt{\frac{3}{2}} \Delta  & 0 \\
 0 & 0 & 0 & 0 & i \sqrt{2} \Delta  & \frac{i \Delta }{\sqrt{2}} & \frac{i \Delta
   }{\sqrt{2}} & i \sqrt{2} \Delta  \\
 0 & 0 & -2 \varepsilon  & 0 & \Delta  & 0 & \Delta  & 0 \\
 0 & 0 & 0 & 2 \varepsilon  & 0 & -\Delta  & 0 & -\Delta  \\
 0 & -i \sqrt{2} \Delta  & \Delta  & 0 & -b t-\varepsilon  & 0 & 0 & 0 \\
 i \sqrt{\frac{3}{2}} \Delta  & -\frac{i \Delta }{\sqrt{2}} & 0 & -\Delta  & 0 &
   \varepsilon -b t & 0 & 0 \\
 i \sqrt{\frac{3}{2}} \Delta  & -\frac{i \Delta }{\sqrt{2}} & \Delta  & 0 & 0 & 0 & b
   t-\varepsilon  & 0 \\
 0 & -i \sqrt{2} \Delta  & 0 & -\Delta  & 0 & 0 & 0 & b t+\varepsilon  \\
\end{array}
\right).
\end{equation}

The respective $\hat A_{ad,bt}$ operator:

\begin{equation}
    \hat A_{ad,bt}= 
   \left(
\begin{array}{cccccccc}
 0 & 0 & 0 & 0 & -\frac{i \sqrt{\frac{3}{2}} \Delta }{b} & \frac{i \sqrt{\frac{3}{2}}
   \Delta }{b} & \frac{i \sqrt{\frac{3}{2}} \Delta }{b} & -\frac{i \sqrt{\frac{3}{2}}
   \Delta }{b} \\
 0 & 0 & \frac{i \sqrt{2} \Delta ^2}{b \varepsilon } & \frac{i \sqrt{2} \Delta ^2}{b
   \varepsilon } & \frac{i \Delta }{\sqrt{2} b} & \frac{i \Delta }{\sqrt{2} b} & \frac{i
   \Delta }{\sqrt{2} b} & \frac{i \Delta }{\sqrt{2} b} \\
 0 & -\frac{i \sqrt{2} \Delta ^2}{b \varepsilon } & -2 t & 0 & -\frac{\Delta }{b} & 0 &
   \frac{\Delta }{b} & 0 \\
 0 & -\frac{i \sqrt{2} \Delta ^2}{b \varepsilon } & 0 & 2 t & 0 & -\frac{\Delta }{b} & 0
   & \frac{\Delta }{b} \\
 \frac{i \sqrt{\frac{3}{2}} \Delta }{b} & -\frac{i \Delta }{\sqrt{2} b} & -\frac{\Delta
   }{b} & 0 & \frac{\Delta ^2-\varepsilon  (b t+\varepsilon )}{b \varepsilon } &
   -\frac{\Delta ^2}{b \varepsilon } & 0 & 0 \\
 -\frac{i \sqrt{\frac{3}{2}} \Delta }{b} & -\frac{i \Delta }{\sqrt{2} b} & 0 &
   -\frac{\Delta }{b} & -\frac{\Delta ^2}{b \varepsilon } & \frac{(\Delta -\varepsilon )
   (\Delta +\varepsilon )}{b \varepsilon }+t & 0 & 0 \\
 -\frac{i \sqrt{\frac{3}{2}} \Delta }{b} & -\frac{i \Delta }{\sqrt{2} b} & \frac{\Delta
   }{b} & 0 & 0 & 0 & -\frac{b \varepsilon  t+\Delta ^2-\varepsilon ^2}{b \varepsilon } &
   \frac{\Delta ^2}{b \varepsilon } \\
 \frac{i \sqrt{\frac{3}{2}} \Delta }{b} & -\frac{i \Delta }{\sqrt{2} b} & 0 &
   \frac{\Delta }{b} & 0 & 0 & \frac{\Delta ^2}{b \varepsilon } & \frac{\varepsilon
   ^2-\Delta ^2}{b \varepsilon }+t \\
\end{array}
\right).
\end{equation}

The solution is similar to what we presented for the 6-dimensional case in the main text.
\end{section}

\end{document}